\documentclass[twocolumn,showpacs,preprintnumbers,amsmath,amssymb]{revtex4-1}

\usepackage{amssymb}
\usepackage{amsmath}
\usepackage{graphicx}
\usepackage{color}
\usepackage{hyperref}

\newcommand{\be}{\begin{equation}}
\newcommand{\ee}{\end{equation}}
\newcommand{\bea}{\begin{eqnarray}}
\newcommand{\eea}{\end{eqnarray}}
\newcommand{\beal}{\begin{aligned}}
\newcommand{\eeal}{\end{aligned}}

\usepackage{color}

\begin{document}



\title{Black hole evaporation in Conformal (Weyl) gravity}


\author{Hao Xu}
\email{xuh5@sustc.edu.cn}
\affiliation{Department of Physics, Southern University of Science and Technology, Shenzhen 518055, China}
\affiliation{Department of Physics, University of Science and Technology of China, Hefei 230026, China}

\author{Man-Hong Yung}
\affiliation{Institute for Quantum Science and Engineering and Department of Physics, Southern University of Science and Technology, Shenzhen 518055, China}
\affiliation{Shenzhen Key Laboratory of Quantum Science and Engineering, Shenzhen, 518055, China}


\date{\today}

\begin{abstract}
In this work, we explore all possible scenarios in the evaporation process of a spherical neutral AdS black hole in four-dimensional conformal gravity, where the equations of states are branched. In one branch, the final states correspond to the extremal black hole where the total decay time is divergent given any initial mass. In the other branch, the black holes always evaporate completely with in a finite time; the total decay time depends linearly on the AdS radius, when the mass is taken to be infinity.

\end{abstract}


\maketitle

Recently, conformal (Weyl) gravity, which is described by a pure Weyl squared action, has attracted a considerable amount of interest as an alternative theory to Einstein gravity. From the equation of motion and the symmetry of conformal gravity, any conformal class of the Einstein solution arises naturally as a solution to the conformal gravity. In particular, subject to the Neumann boundary condition, conformal gravity can single out Einstein's solution~\cite{Maldacena:2011mk,Anastasiou:2016jix}. Moreover, unlike Einstein's gravity, conformal gravity has been shown to be perturbatively renormalizable in four dimensions~\cite{Mannheim:2009qi}, leading to many interesting alternative approaches to quantum gravity \cite{Stelle:1976gc}.

Another appealing aspect of conformal gravity comes from cosmology. Although Einstein's gravity can describe the physics within the scale of the Solar system perfectly, there are still many unsolved puzzles when applied to a larger scale, such as the inconsistency with the observation of galactic rotation curves and accelerating universe. Consequently, unknown entities namely ``dark matter and dark energy", have to be introduced in order to fix the inconsistency problems. Thus, one may wonder whether it is possible to modify the nature of gravity to explain the physics at a larger scale, while maintaining the correct behaviors at the scale of the Solar system. Since conformal gravity admits more solutions than Einstein's gravity, it can produce the effective potential consistent with the observed phenomenon, making it a compelling modified gravity theory~\cite{Mannheim:1988dj,Mannheim:2005bfa,Mannheim:2010ti,Mannheim:2011ds}.

Although conformal gravity and Einstein's gravity may share the same space-time solutions, thermodynamic quantities of the black hole, such as the entropy and mass, depend on the action rather than the metric. The difference in the actions means that the black-hole thermodynamics of the two gravity theories would be different. The thermodynamic phase structure for the conformal gravity in the four-dimensional AdS space-time has been explored~\cite{Lu:2012xu,Xu:2014kwa,Xu:2017ahm,Xu:2018vxf}; the phase structure of the conformal gravity includes two branches of equations of states, zeroth-order phase transition, and Hawking-Page-like phase transition.

However, there are still some puzzles left unsolved. More than forty years after the discovery of Hawking radiation, its precise nature and consequences are still, to a large extent, open questions. The radiation causes a black hole to lose mass and thermal entropy, if there is not sufficient incoming radiation to balance it. For a non-rotating neutral Schwarzschild black hole with mass $M_0$ in four-dimensional asymptotically flat space-time without incoming radiation, we expect it has a lifetime $t\sim M_0^3$, which grows as the cube of the mass and diverges when $M_0\rightarrow +\infty$ \cite{Page:1976df,Page:2004xp}. Similar analysis was also carried out for a spherical black hole in AdS space-time \cite{Page:2015rxa} and was later extended to flat and hyperbolic black holes~\cite{Ong:2015fha}. The gravitational potential of AdS space-time on the boundary acts like a finite confining box for the black hole. Massless radiation particles can escape to infinity and reflect back in a finite time. However, if one chooses an absorbing boundary condition \cite{Avis:1977yn} so that Hawking radiation is absorbed by the AdS boundary preventing the returning of the radiation, then one can calculate the time for the black-hole mass to drop from one initial value to another, which is of great interest in holography \cite{Maldacena:1997re,Witten:1998qj}. In particular, Page made a counter-intuitive discovery~\cite{Page:2015rxa} that this time does not diverge even when the initial black hole mass is taken to be infinity.

Given the thermodynamic phase structure of conformal gravity in AdS space-time, it is natural to investigate the black-hole evaporation process, which points to an area where extensive research should be carried out. In particular, we are interested in answering the question, whether the total decay time is also finite when the black hole mass is taken to infinity.

We start by giving a brief review of the black hole thermodynamics in conformal gravity \cite{Lu:2012xu,Xu:2014kwa,Xu:2017ahm,Xu:2018vxf}. The gravity action is proportional to the square of the Weyl tensor,
\begin{align}
S=\alpha\int \mathrm{d}^4x \sqrt{-g}C^{\mu\nu\rho\sigma}
C_{\mu\nu\rho\sigma} \ ,
\label{action}
\end{align}
where the coupling constant $\alpha$ is associated with a dimension of [length]$^2$. Notice that some thermodynamic quantities, such as the black hole mass, entropy, and Gibbs free energy, are proportional to $\alpha$. However, the exact value of $\alpha$ does not change any qualitative features of the thermodynamics. Without loss of generality, we set $\alpha=1$ in the present work. In other words, the black hole thermodynamic quantities are described in unit of $\alpha$.

The simplest form of spherical black hole solution to four-dimensional conformal gravity \cite{Riegert:1984zz,Klemm:1998kf} is
\begin{align}
  &\mathrm{d} s^2=-f(r)\mathrm{d}t^2+\frac{\mathrm{d}r^2}{f(r)}+r^2(\mathrm{d}\theta^2+\sin^2 \theta\mathrm{d}\phi^2),
  \label{metric}
\end{align}
where
\begin{align}
  f(r)=c_0+c_1 r+\frac{d}{r}-\frac{1}{3}\Lambda r^2.
\end{align}
Due to the conformal symmetry, the Weyl rescaling of the  metric above remains a static and spherically-symmetric solution. Here the four integral constants, $c_0, c_1, d, \Lambda$, are subject to the following constraint:
\begin{align}
  c_0^2=3c_1d+1,
\label{relation}
\end{align}
which implies that we have a discrete freedom in choosing the constant $c_0$,
\begin{align}
c_0=\pm\sqrt{3c_1d+1}.
\label{c0}
\end{align}
If we take $c_1=0$ and $c_0=1$, the solution reduces to the well-known Schwarzschild (A)dS black hole.

Here the $\Lambda$ plays the role of ``cosmological constant", which comes from the integral instead of the action. Furthermore, we adopt the language of extended phase space and treat it as effective pressure $P=-\frac{\Lambda}{8\pi}$, which is allowed to vary\cite{Gibbons:1996af,Kastor:2010gq,Xu:2013zea,Anabalon:2015xvl,
Astefanesei:2018vga,Kastor:2009wy,Kubiznak:2012wp,Mann:2015luq,
Gregory:2017sor,Gunasekaran:2012dq,Cai:2013qga,
Altamirano:2013uqa,Zou:2013owa,Altamirano:2014tva,Wei:2014hba,
Johnson:2014yja,Xu:2014tja,Frassino:2014pha,
Dolan:2014vba,Xu:2015rfa,Xu:2015hba,Liu:2016uyd,Majhi:2016txt,Ma:2016vop,Zou:2016sab,Bhattacharya:2017hfj,Zou:2017juz,
Bhattacharya:2017nru,Xu:2018fag,Liu:2018jld}.

The black hole mass is defined to be the conserved charge of the time-like killing vector~\cite{Lu:2012xu}
\begin{align}
  M=\,{\frac {\left( c_{{1}}c_{{0}}-c_{{1}}-16\pi P\,d
 \right) }{12\pi }}.
 \label{mass}
\end{align}
The temperature is proportional to the surface gravity at the black hole horizon $r_0$
\begin{align}
  T=\,{\frac {8\pi P\,r_0^{3}-3\,c_{{0}}r_{{0}}-6\,d}{12\pi \,{r
_{{0}}}^{2}}},
\label{temperature}
\end{align}
and its conjugate, i.e. the entropy is
\begin{align}
  S=\,{\frac {\,\left( r_0-c_{{0}}r_{{0}}-\,  3\,
d \right) }{3r_{{0}}}}.
\label{S}
\end{align}
Note that the entropy depends on a combination of $c_0$, $r_0$ and $d$, rather than just being proportional to the area of the horizon \cite{Lu:2012xu}. Furthermore, the first law of black hole thermodynamics and the Smarr formula in extended phase space have been investigated in \cite{Lu:2012xu,Xu:2014kwa,Xu:2017ahm,Xu:2018vxf} and will not be covered here.

Now, applying $f(r_0)=0$ and \eqref{relation} to eliminate other unnecessary coefficients, we found that the equations of states are branched~\cite{Xu:2017ahm}. One set of the solutions is given by,
\begin{align}
\begin{split}
  &T_1=2Pr_0-\frac{\sqrt{-32\pi Pc_1r_0^3-3c_1^2r_0^2+4}}{8\pi r_0}+\frac{c_1}{8\pi},\\
  &S_1=\frac{8}{3}\pi P r_0^2-\frac{1}{3}\sqrt{-32\pi Pc_1r_0^3-3c_1^2r_0^2+4}+\frac{1}{3}.
\label{solution1}
\end{split}
\end{align}
The other one is given by,
\begin{align}
\begin{split}
  &T_2=2Pr_0+\frac{\sqrt{-32\pi Pc_1r_0^3-3c_1^2r_0^2+4}}{8\pi r_0}+\frac{c_1}{8\pi},\\
  &S_2=\frac{8}{3}\pi P r_0^2+\frac{1}{3}\sqrt{-32\pi Pc_1r_0^3-3c_1^2r_0^2+4}+\frac{1}{3}.
\label{solution2}
\end{split}
\end{align}
Here we use $r_0$ as an intermediate parameter to avoid the overwhelming complexity of formula $T(P,S)$.

Next, we investigate the possible critical behavior \cite{Xu:2014kwa,Xu:2017ahm} by considering the following two conditions:
\begin{align}
\frac{\partial T}{\partial S}=0,\quad \frac{\partial^2 T}{\partial S^2}=0,
\end{align}
yielding the critical point,
\begin{align}
\begin{split}
&P_c=\frac{(2-3X^2-\sqrt{3}X)c_1^2}{24\pi X^3},\quad T_c=\frac{c_1}{6\pi X^2},\\
&S_c=\frac{\sqrt{3}X+3X+2}{9X},
\end{split}
\end{align}
where $X=\sqrt{2}-2\sqrt{3}/{3}$. This critical point only appears for $c_1>0$ and corresponds to the zeroth-order phase transition.

Now we concentrate on the black hole evaporation process. The black hole mass $M$ and entropy $S$ should be monotonically-decreasing functions of time. We shall apply the geometrical optics approximation, which assumes massless quanta move along null geodesics \cite{Page:2015rxa}. If we orient the angular coordinates and normalize the affine parameter $\lambda$, the geodesic equation takes the form
\begin{align}
\bigg(\frac{\mathrm{d}r}{\mathrm{d}\lambda}\bigg)^2=E^2-J^2\frac{f(r)}{r^2},
\end{align}
where $E=f(r)\frac{\mathrm{d}t}{\mathrm{d}\lambda}$ and $J=r^2\frac{\mathrm{d}\theta}{\mathrm{d}\lambda}$ are the energy and angular momentum of the massless quanta respectively. Considering a null geodesic comes from the black hole horizon, it will return to the hole when there is a turning point satisfying $\big(\frac{\mathrm{d}r}{\mathrm{d}\lambda}\big)^2=0$ thus cannot be detected by the observer on the AdS boundary. If we introduce $b\equiv\frac{J}{E}$, then the massless quanta can reach infinity if and only if $b$ satisfies
\begin{align}
\frac{1}{b^2}\geq \frac{f(r)}{r^2}
\end{align}
for \emph{any} $r\geq r_0$. Therefore, we need to find the maximal value of $\frac{f(r)}{r^2}$. Depending on the exact form of $f(r)$, there are different cases. The $\frac{f(r)}{r^2}$ may admit a maximal value corresponding to an unstable photon orbit $r_p$, such as in the spherical AdS black hole case \cite{Page:2015rxa}. Or it may be monotonically increasing and approaches $\frac{8\pi P}{3}$ near the boundary, such as in the flat AdS black hole case \cite{Ong:2015fha}.

For the conformal gravity with metric \eqref{metric}, we can investigate the $\frac{f(r)}{r^2}$ for different signs of $c_1$ and $c_0$. In TABLE \ref{tab1}, we list all possible values of $r_p$ so that we can find the impact factor $b_c^2=\frac{r_p^2}{f(r_p)}$. Here the $none$ means the $\frac{f(r)}{r^2}$ is a monotonic function of $r_0$ and the impact factor $b_c^2=\frac{3}{8\pi P}$.
\begin{table}[!htbp]
\centering
\begin{tabular}{|c|c|c|c|c|}
\hline
$$ ~&~  $c_1=0$ ~&~   $c_1>0$ ~&~ $c_1<0$\\
\hline
$c_0>0$ ~&~ $-\frac{3d}{2}$ ~&~ $\frac{1-\sqrt{1+3c_1 d}}{c_1}$ ~&~ $\frac{1-\sqrt{1+3c_1 d}}{c_1}$   \\
\hline
$c_0<0$ ~&~ $none$ ~&~ $\frac{1+\sqrt{1+3c_1 d}}{c_1}$ ~&~ $none$  \\
\hline
\end{tabular}
\caption{The value of the photon orbit radius $r_p$.}
\label{tab1}
\end{table}

When $c_1=0$ and $c_0>0$, the black hole metric reduces to the Schwarzschild AdS metric, and the photon orbit radius $r_p=-\frac{3d}{2}$. When $c_1\neq 0$ and $c_0>0$,
\begin{align}
\begin{split}
r_p&=\frac{1-\sqrt{1+3c_1 d}}{c_1}\\
   &\sim-\frac{3d}{2}(1-\frac{3d}{4}c_1+\frac{9d^2}{8}c_1^2+O(d^3c_1^3)),
\end{split}
\end{align}
which goes to $-\frac{3d}{2}$ as $c_1\rightarrow 0$. When $c_1>0$ and $c_0<0$,
\begin{align}
\begin{split}
r_p&=\frac{1+\sqrt{1+3c_1 d}}{c_1}\\
   &\sim\frac{2}{c_1}+\frac{3d}{2}-\frac{9d^2}{8}c_1+\frac{27d^3}{16}c_1^2+O(d^4c_1^3),
\end{split}
\end{align}
which goes to $+\infty$ as $c_1\rightarrow 0$. In $c_1\leq 0$ and $c_0<0$, the $\frac{f(r)}{r^2}$ is a monotonic function of $r_0$.\\

According to the Stefan-Boltzmann law, we conclude that in four dimensional space-time, the Hawking emission power (luminosity) is
\begin{align}
\frac{\mathrm{d} M}{\mathrm{d}t}=-C b_c^2 T^4
\end{align}
with some numerical constants $C$. Since we are only concerned about the qualitative features of the evaporation process, we shall ignore this constant. The qualitative features of the thermodynamic phase structure and the impact factors depend on the signs of $c_1$ and $c_0$. We will investigate them in all different situations.
\\

\begin{figure}
\begin{center}
\includegraphics[width=0.3\textwidth]{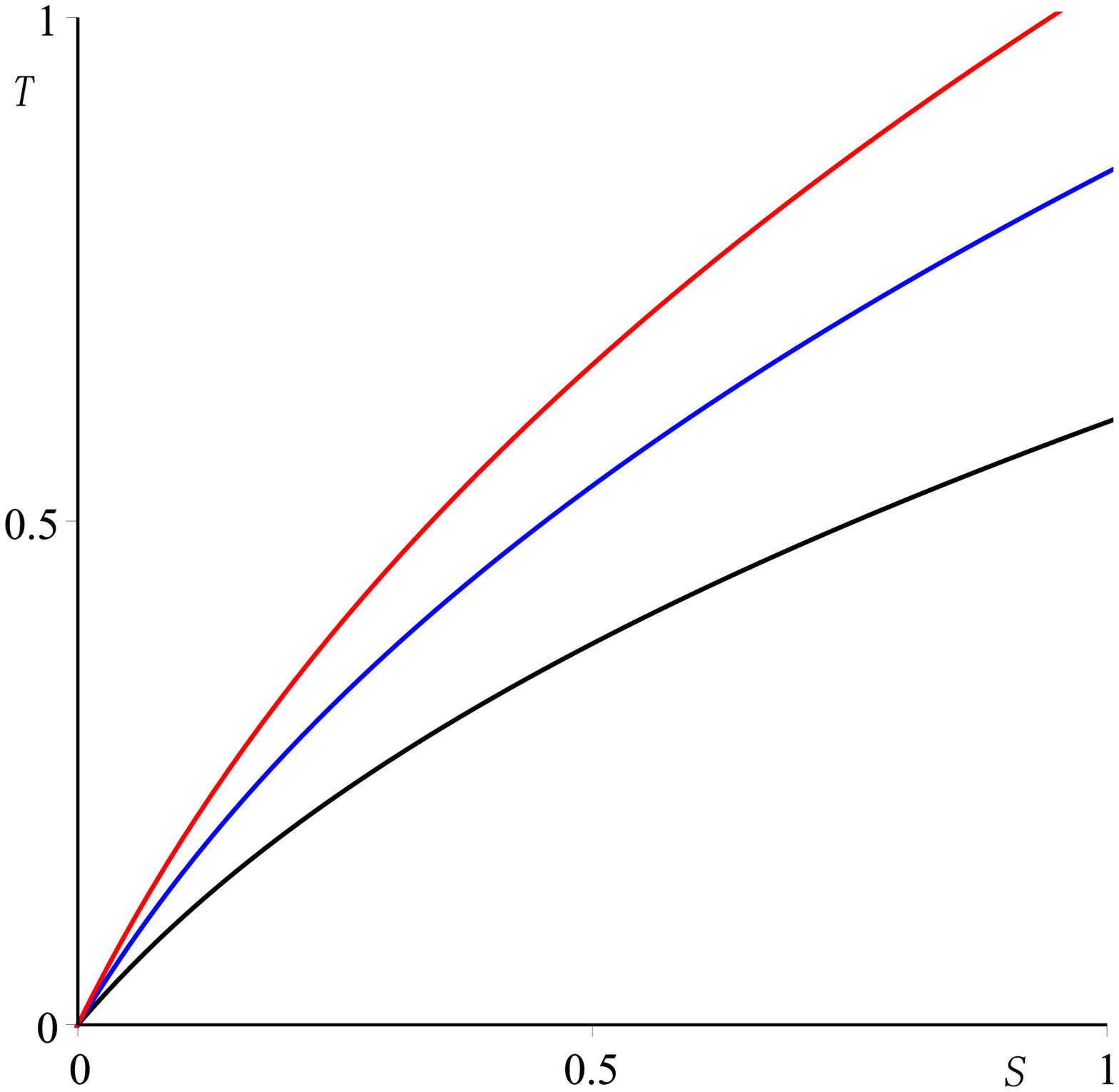}
\includegraphics[width=0.3\textwidth]{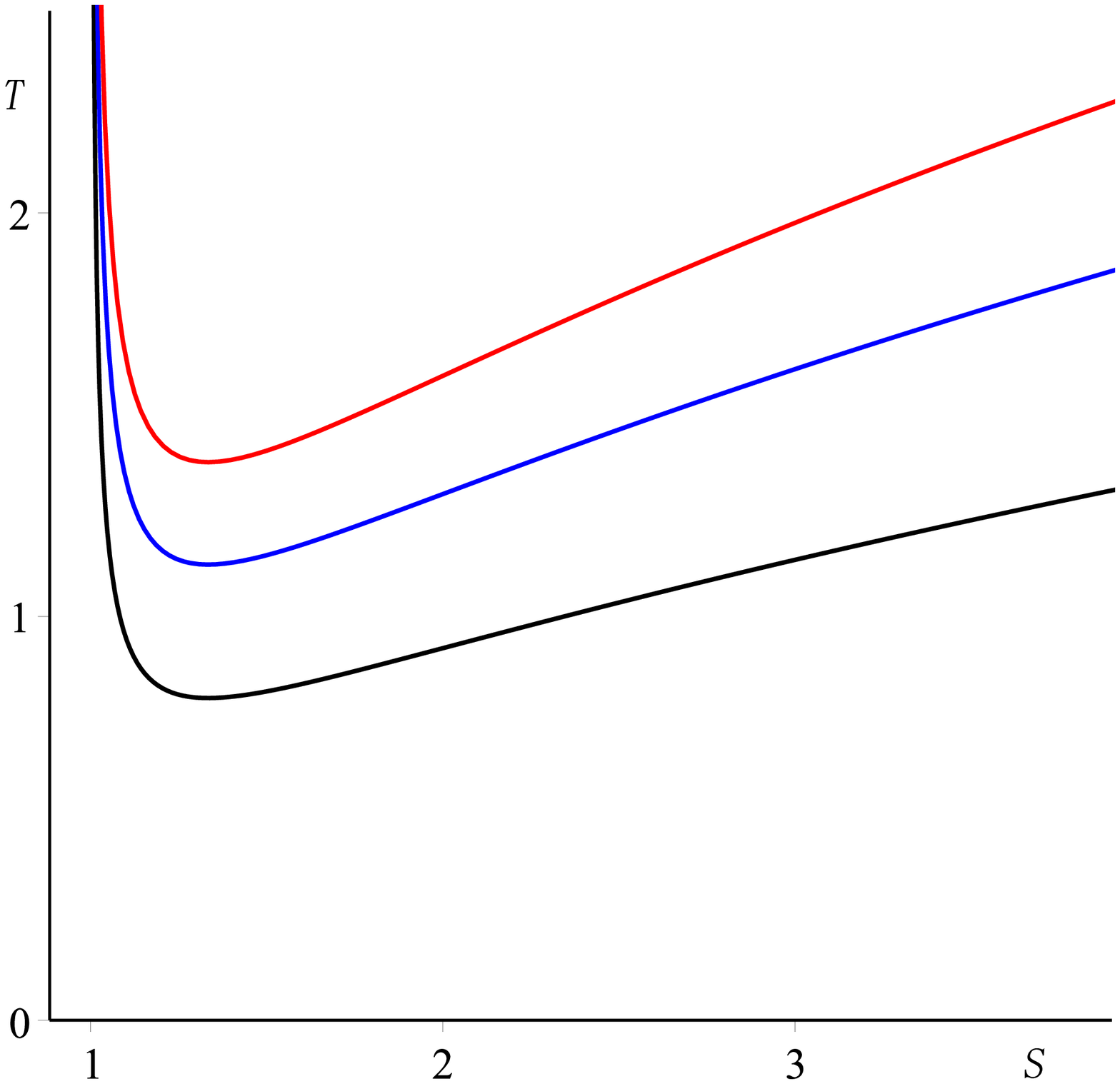}
\caption{The isobaric curves in $T-S$ planes for $c_1=0$. The two figures correspond to $c_0=-1$ and $c_0=1$ respectively. In each figure the pressure $P$ corresponds to $P=1, 2, 3$ from bottom to top.}
\label{fig1}
\end{center}
\end{figure}

1. $c_1=0$. In this case, the equations of states for the two solutions become
\begin{align}
  T_1=2Pr_0-\frac{1}{4\pi r_0},\quad S_1=\frac{8}{3}\pi P r_0^2-\frac{1}{3},
\end{align}
and
\begin{align}
  T_2=2Pr_0+\frac{1}{4\pi r_0},\quad S_2=\frac{8}{3}\pi P r_0^2+1,
\end{align}
which correspond to $c_0=-1$ and $c_0=1$ respectively. The isobaric curves are presented in FIG.\ref{fig1}. For $c_0=-1$ (the first figure in in FIG.\ref{fig1}), the curves are monotonically increasing, and $T\rightarrow 0$ as $S\rightarrow 0$, which corresponds to an extremal black hole. The radius of the extremal black hole reads
\begin{equation}
r_0=\frac{1}{\sqrt{8\pi P}}.
\end{equation}
When the black hole evaporates and loses mass and entropy, the temperature decreases so the evaporation process is increasingly difficult.

For $c_0=1$ (the second figure in in FIG.\ref{fig1}), the black hole is in an analogy with the Hawking-Page phase transition in Schwarzschild AdS black hole from Einstein gravity \cite{Hawking:1982dh}. For each $P$ the black hole admits a minimal temperature. The curve with negative slope to the left of the minimum corresponds to small black holes that are thermodynamically unstable. Black hole that reaches these states will soon be evaporated completely. Notice that the black hole entropy includes a pure constant $1$ even in $r_0=0$. It can be removed by introducing a Gauss-Bonnet term in the action \cite{Lu:2012xu}.

\begin{figure}
\begin{center}
\includegraphics[width=0.35\textwidth]{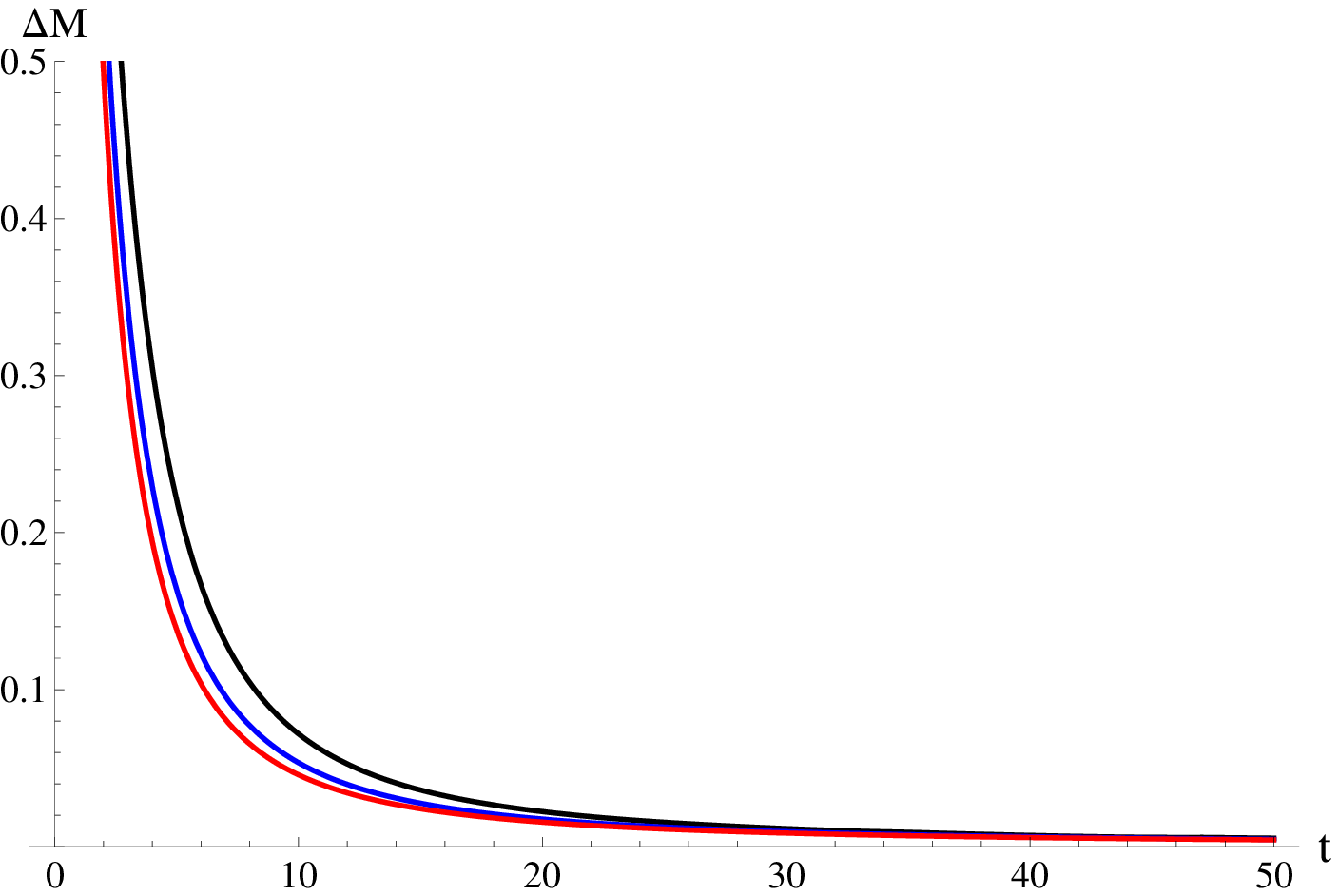}
\includegraphics[width=0.35\textwidth]{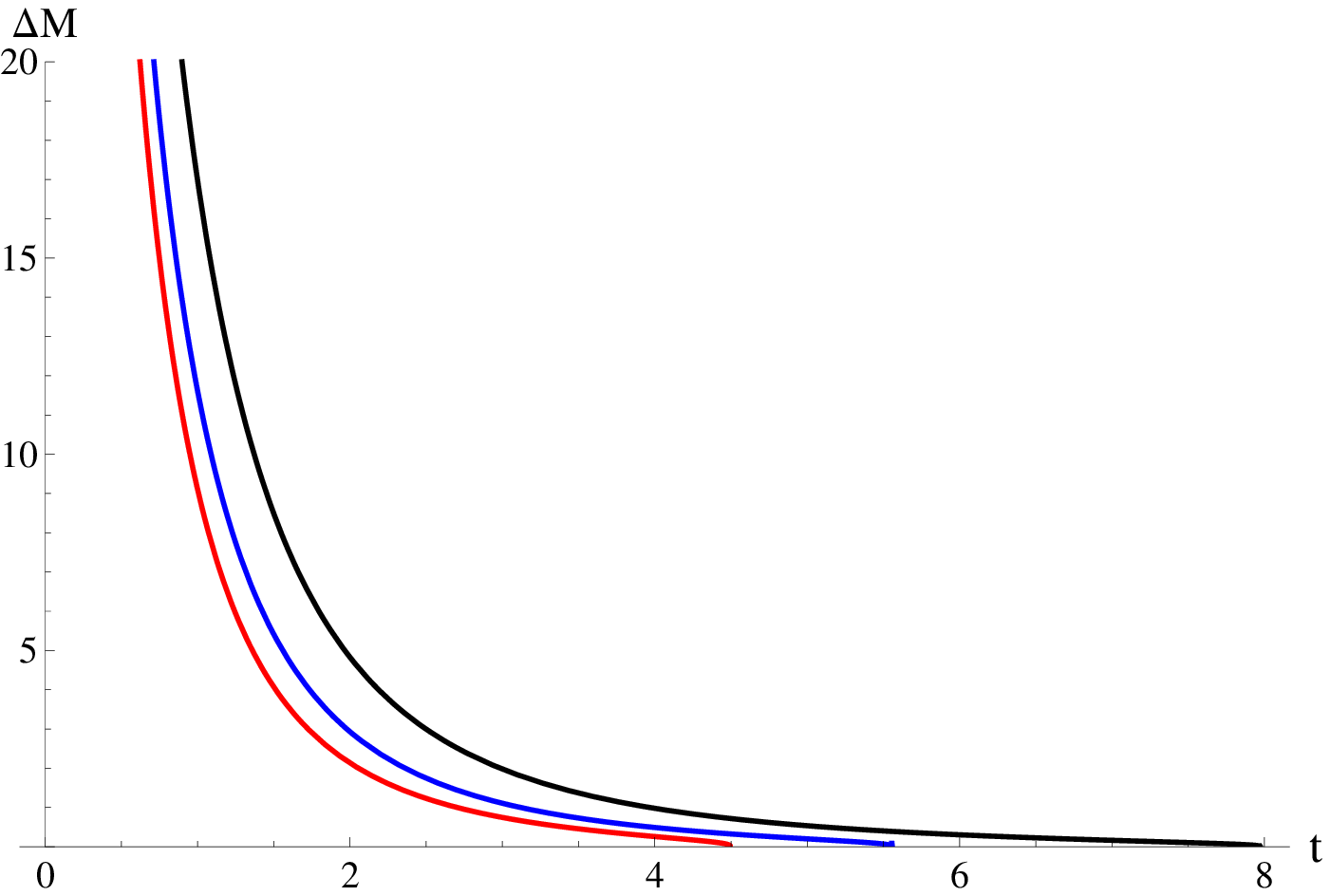}
\caption{The evolution of the black hole in $c_1=0$. The two figures correspond to $c_0=-1$ and $c_0=1$ respectively. In each figure the pressure $P$ corresponds to $P=1, 2, 3$ from top to bottom.}
\label{fig2}
\end{center}
\end{figure}

In FIG.\ref{fig2} we provide some numerical examples of the black hole evolution. We introduce $\Delta M$, defined as $\Delta M=M_0-M_f$. $M_0$ is the initial mass. $M_f$ is the final mass, which corresponds to the $T\rightarrow 0$ in $c_0=-1$ and $T\rightarrow \infty$ on the left side of the minimum in $c_0=1$. In the first figure, we can observe the black hole mass decreases rapidly in the beginning but the evaporation slows down quickly and it will take infinite time to reach zero temperature, which obeys the third law of thermodynamics. However, in the second figure, although we set the initial mass to be very large, the total decay time to Hawking emission is still \emph{finite}. This result agrees with the case of the Schwarzschild AdS black hole in Einstein gravity.

\begin{figure}
\begin{center}
\includegraphics[width=0.35\textwidth]{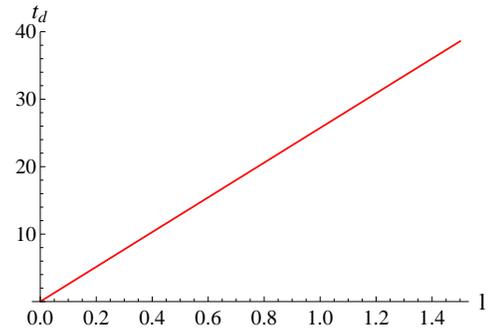}
\caption{The relationship between the total decay time $t_d$ and AdS radius $\ell=\sqrt{\frac{3}{8\pi P}}$ in $c_1=0$ and $c_0=1$.}
\label{fig3}
\end{center}
\end{figure}

It is worth emphasizing that from FIG.\ref{fig2} we can observe the smaller pressure $P$ corresponds the larger total decay time. In \cite{Page:2015rxa} Page proved in four dimensional AdS space-time the decay time is bounded by a time of the order of $\ell^3$, where $\ell=\sqrt{\frac{3}{8\pi P}}$ is the AdS radius. However, although the temperature $T$ in $c_1=0$ and $c_0=1$ shares the same formula with the Schwarzschild AdS black hole, the black hole mass \eqref{mass} is defined as $M=-\frac{4\pi Pd}{3}\sim -\frac{d}{\ell^2}$ (the scaling can be compensated by the coupling constant $\alpha$ in the action), rather than $-d$. We can check that the decay time in conformal gravity is in the \emph{linear} order of $\ell$, not $\ell^3$. See FIG.\ref{fig3} for numerical result.
\\

2. $c_1>0$. The exact value of $c_1$ does not change the qualitative features of the thermodynamic phase structure. Since the thermodynamic quantities are described as functions of $P$ and $r_0$, we can introduce a reduced parameter space $\tilde{r}_0=r_0 c_1$ and $\tilde{P}$$=\frac{P}{c_1^2}$, so that all the thermodynamic quantities, such as $\tilde{M}=\frac{M}{c_1}$ and $\tilde{T}=\frac{T}{c_1}$, have [length]$^0$(constant) scaling. The $c_1$ is used as a unit to describe the quantities. Now the equations of states become
\begin{align}
\begin{split}
  &\tilde{T_1}=2\tilde{P}\tilde{r_0}-\frac{\sqrt{-32\pi \tilde{P}\tilde{r}_0^3-3\tilde{r}_0^2+4}}{8\pi \tilde{r}_0}+\frac{1}{8\pi},\\
  &S_1=\frac{8}{3}\pi \tilde{P} \tilde{r}_0^2-\frac{1}{3}\sqrt{-32\pi \tilde{P}\tilde{r}_0^3-3\tilde{r}_0^2+4}+\frac{1}{3},
\label{solution1}
\end{split}
\end{align}
and
\begin{align}
\begin{split}
  &\tilde{T_2}=2\tilde{P}\tilde{r_0}+\frac{\sqrt{-32\pi \tilde{P}\tilde{r}_0^3-3\tilde{r}_0^2+4}}{8\pi \tilde{r}_0}+\frac{1}{8\pi},\\
  &S_2=\frac{8}{3}\pi \tilde{P} \tilde{r}_0^2+\frac{1}{3}\sqrt{-32\pi \tilde{P}\tilde{r}_0^3-3\tilde{r}_0^2+4}+\frac{1}{3}.
\label{solution2}
\end{split}
\end{align}

\begin{figure}
\begin{center}
\includegraphics[width=0.4\textwidth]{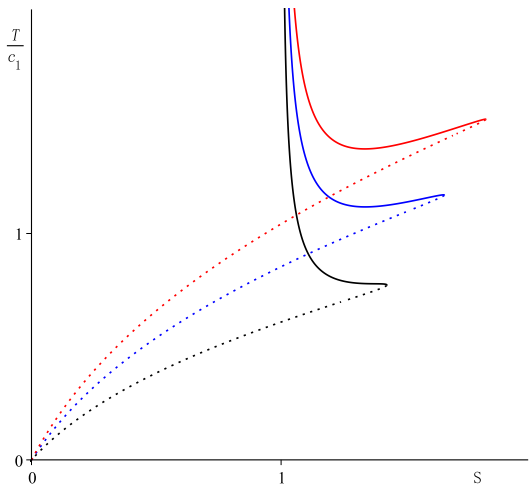}
\caption{The isobaric curves in $\frac{T}{c_1}-S$ planes for $c_1>0$. The dotted and solid line correspond to $c_0<0$ and $c_0>0$ respectively. The pressure $P$ corresponds to $P=1, 2, 3$ from bottom to top.}
\label{fig4}
\end{center}
\end{figure}

In FIG.\ref{fig4} we present the isobaric curves in the $\tilde{T}-S$ plane. The solid and dotted lines correspond to the $c_0>0$ and $c_0<0$ cases respectively. Notice that there is also a region of $(\tilde{T_2},S_2)$ with negative $c_0$. See \cite{Xu:2017ahm} for detailed discussion. We can observe that the radius (and also mass and entropy) is bounded from above for each $P$ due to the $\sqrt{-32\pi \tilde{P}\tilde{r}_0^3-3\tilde{r}_0^2+4}$ term.

Solving the equations of $\tilde{T_1}=0$ and $S_1=0$, we can find there is also an extremal black hole for each $P$ in $c_0<0$. The radius of the extremal black hole reads
\begin{equation}
\tilde{r}_0=\frac{\sqrt{32\pi P+1}-1}{16\pi P}.
\end{equation}
For $c_0>0$, the temperature $\tilde{T_2}\rightarrow \infty$ as $\tilde{r_0}\rightarrow 0$.

We give a brief discussion about the black hole evaporation in AdS space-time. Unlike asymptotically flat space-time, in AdS space-time the black hole mass is determined by both the cosmological constant and the black hole radius, thus it seems unnatural to compared the black hole evaporation process in different cosmological constants. The best we can do is to take the limit $r_0\rightarrow \infty$ and then study the relationship between the total decay time and the cosmological constant. In Einstein gravity, the total decay time is in the order of $\ell^3$, while in conformal gravity with $c_1=0$ and $c_0=1$, it is in the order of $\ell$. However, in $c_1>0$ the black hole mass and radius are bounded from above for each $P$, so we cannot take the limit $r_0\rightarrow \infty$. The numeral analysis shows in $c_0>0$ the black hole can evaporate completely in finite time, depending on the explicit values of the $P$ and initial mass. In $c_0<0$ the total decay time is divergent in any $P$ and any initial black hole radius. We omit the numerical example.

3. $c_1<0$. In this case, similarly we can also introduce the dimensionless quantities $\tilde{r}_0=r_0 |c_1|$ and $\tilde{P}$$=\frac{P}{c_1^2}$, and the black hole mass $\tilde{M}=\frac{M}{|c_1|}$ and temperature $\tilde{T}=\frac{T}{|c_1|}$. The equations of states now take the form
\begin{align}
\begin{split}
  &\tilde{T_1}=2\tilde{P}\tilde{r_0}-\frac{\sqrt{32\pi \tilde{P}\tilde{r}_0^3-3\tilde{r}_0^2+4}}{8\pi \tilde{r}_0}-\frac{1}{8\pi},\\
  &S_1=\frac{8}{3}\pi \tilde{P} \tilde{r}_0^2-\frac{1}{3}\sqrt{32\pi \tilde{P}\tilde{r}_0^3-3\tilde{r}_0^2+4}+\frac{1}{3},
\label{solution1}
\end{split}
\end{align}
and
\begin{align}
\begin{split}
  &\tilde{T_2}=2\tilde{P}\tilde{r_0}+\frac{\sqrt{32\pi \tilde{P}\tilde{r}_0^3-3\tilde{r}_0^2+4}}{8\pi \tilde{r}_0}-\frac{1}{8\pi},\\
  &S_2=\frac{8}{3}\pi \tilde{P} \tilde{r}_0^2+\frac{1}{3}\sqrt{32\pi \tilde{P}\tilde{r}_0^3-3\tilde{r}_0^2+4}+\frac{1}{3}.
\label{solution2}
\end{split}
\end{align}

\begin{figure}
\begin{center}
\includegraphics[width=0.3\textwidth]{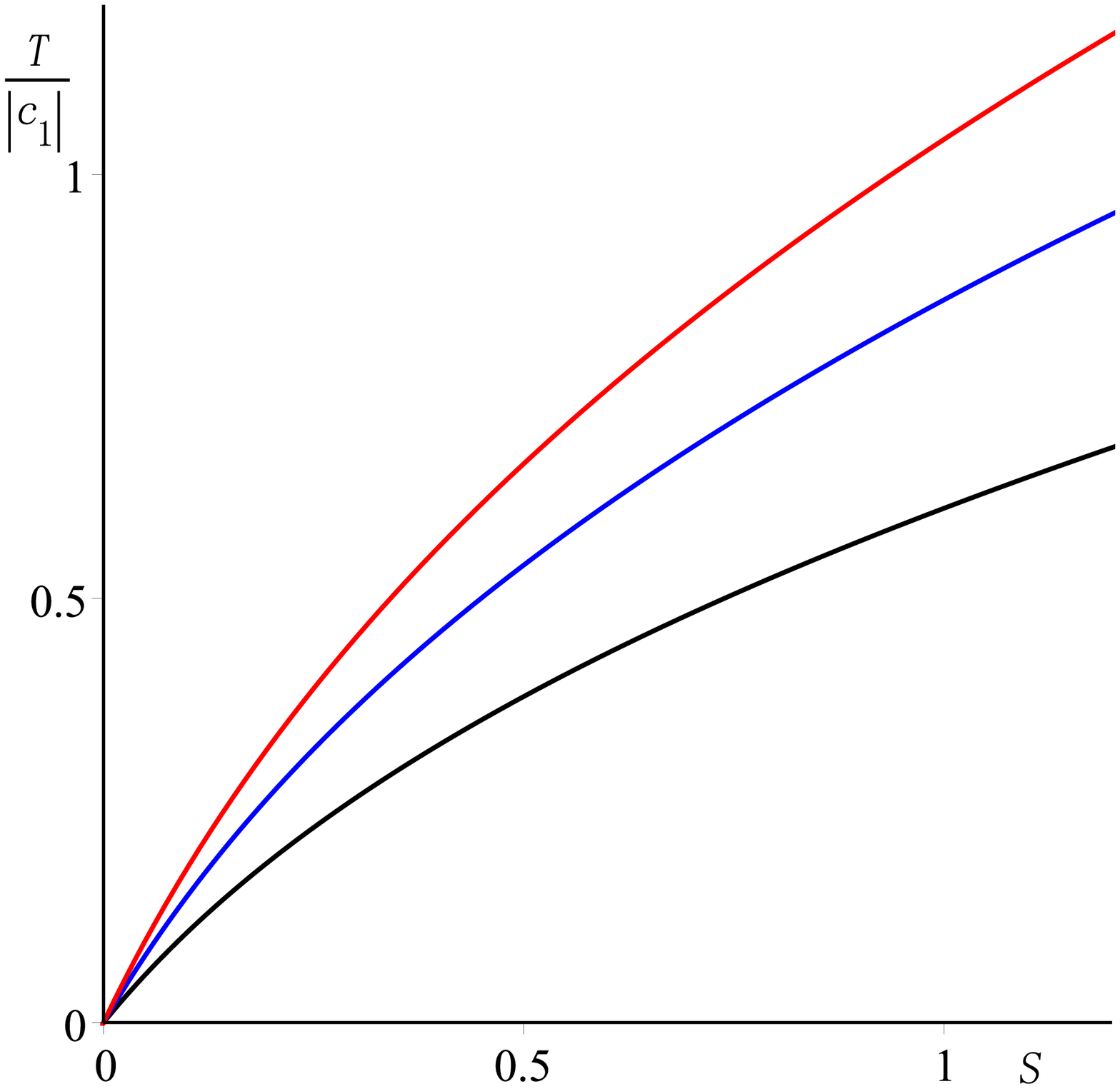}
\includegraphics[width=0.3\textwidth]{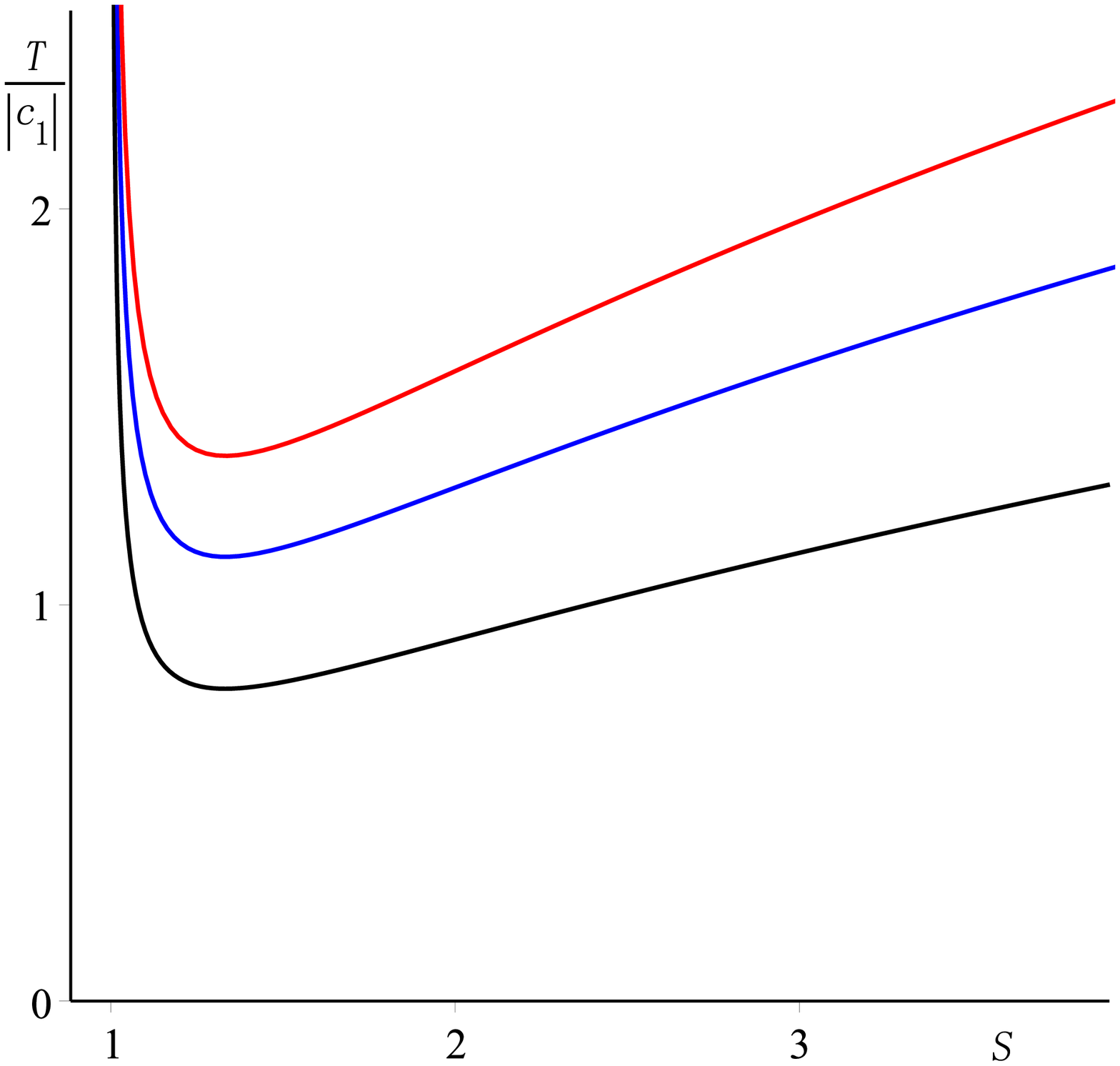}
\caption{The isobaric curves in $\frac{T}{|c_1|}-S$ planes for $c_1<0$. The two figures correspond to $c_0<0$ and $c_0>0$ respectively. In each figure the pressure $P$ corresponds to $P=1, 2, 3$ from bottom to top.}
\label{fig5}
\end{center}
\end{figure}

\begin{figure}
\begin{center}
\includegraphics[width=0.35\textwidth]{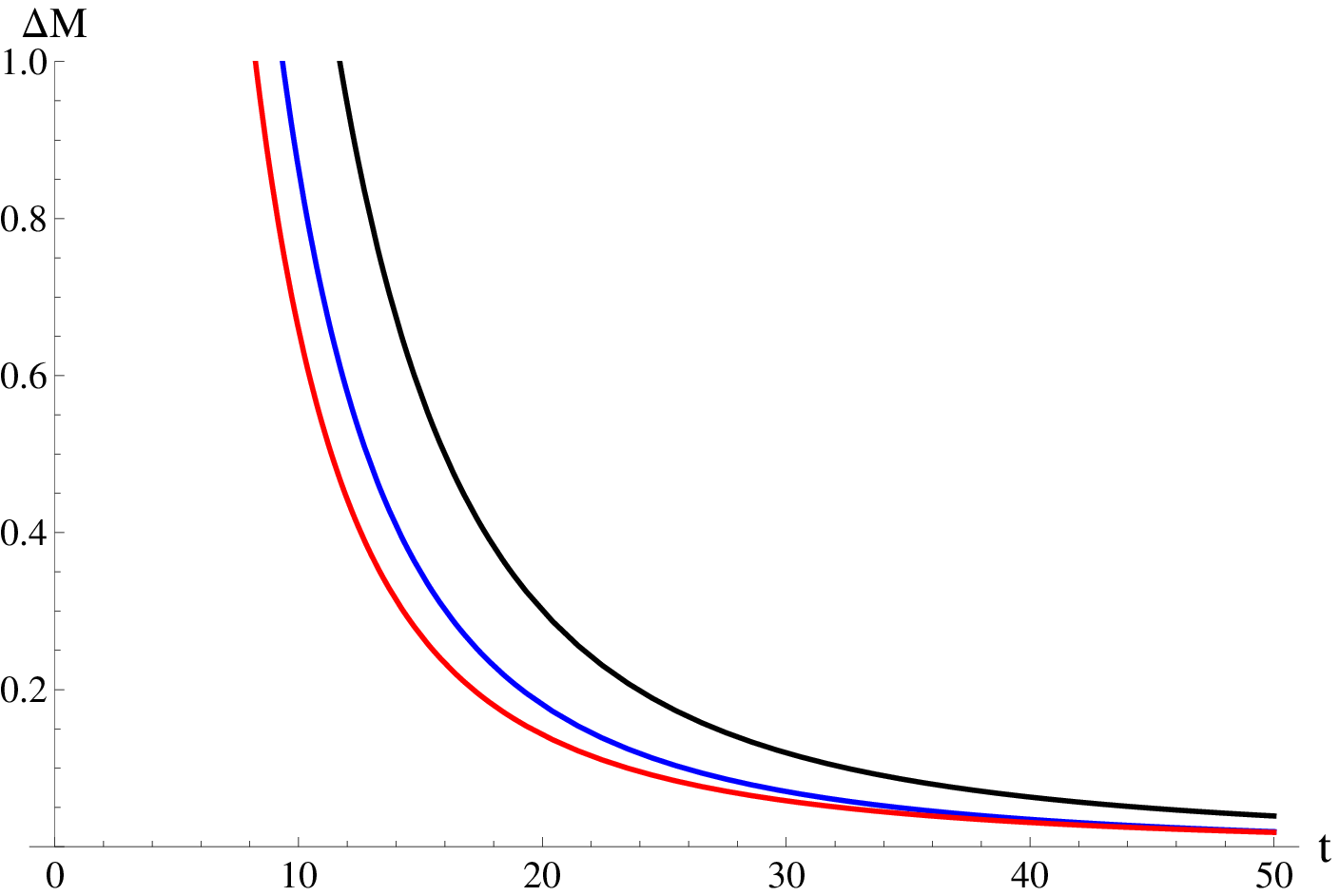}
\includegraphics[width=0.35\textwidth]{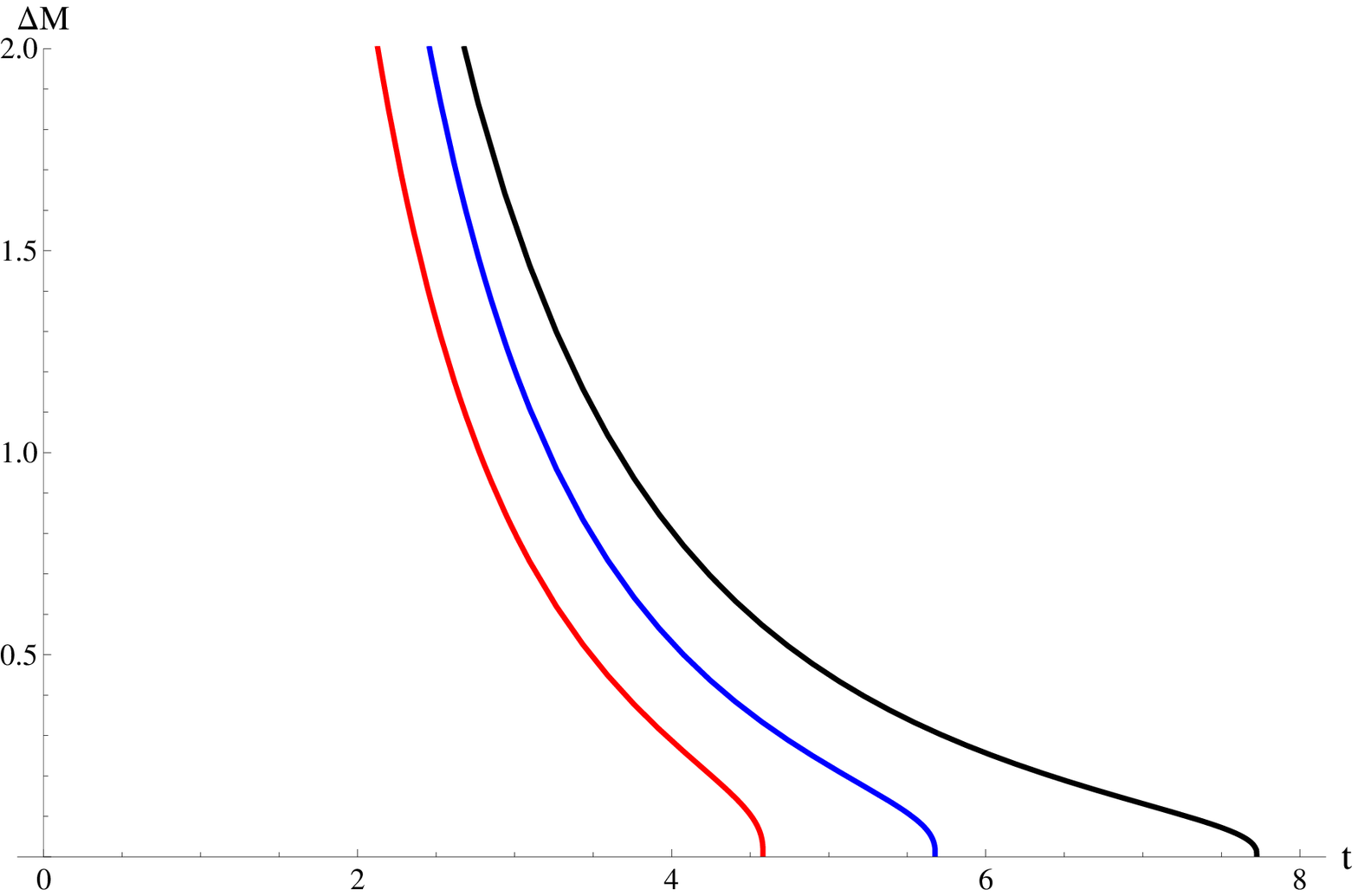}
\caption{The evolution of the black hole in $c_1<0$. The two figures correspond to $c_0<0$ and $c_0>0$ respectively. In each figure the pressure $P$ corresponds to $P=1, 2, 3$ from top to bottom.}
\label{fig6}
\end{center}
\end{figure}

\begin{figure}
\begin{center}
\includegraphics[width=0.35\textwidth]{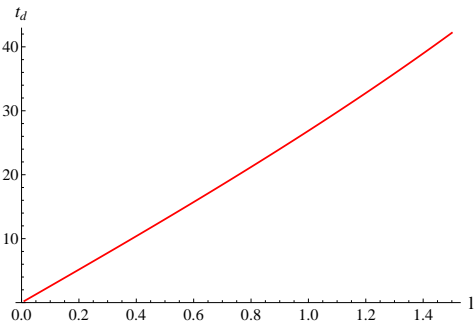}
\caption{The relationship between the total decay time and AdS radius $\ell=\sqrt{\frac{3}{8\pi P}}$ in $c_1<0$ and $c_0>0$.}
\label{fig7}
\end{center}
\end{figure}

Now the black hole radius and mass are unbound. In FIG.\ref{fig5} we can observe the the isobaric curves are in an analogy the the case of $c_1=0$. For $c_0<0$ (the first figure in FIG.\ref{fig5}), we can find there is also an extremal black hole for each $P$ whose radius reads
\begin{equation}
\tilde{r}_0=\frac{\sqrt{32\pi P+1}+1}{16\pi P}.
\end{equation}
For $c_0>0$ (the second figure in FIG.\ref{fig5}) the black hole is in an analogy with the the Schwarzschild AdS black hole. The $c_1$ which acts as an correction can be ignored in large mass limit, thus does not change the qualitative features. In FIG.\ref{fig6} we present some numerical examples of the black hole evolution. In $c_0<0$ the total decay time is divergent and in $c>0$ it reminds finite. The relationship between the total decay time and AdS radius is presented in FIG.\ref{fig7}.
\\

In this paper we explore the spherical neutral AdS black hole evaporation process in four dimensional conformal gravity. We investigate the system in three cases: $c_1=0$, $c_1>0$ and $c_1<0$. In each case we have two branches of equations of states, which corresponds to $c_0<0$ and $c_0>0$. For $c_0<0$, the equations of states admit an extremal black hole with zero temperature and entropy in all three cases. In $c_1=0$ the extremal black hole radius $r_0=\frac{1}{\sqrt{8\pi P}}$, while in $c_1>0$ the $\tilde{r}_0=\frac{\sqrt{32\pi P-1}+1}{16\pi P}$, and in $c_1<0$ the $\tilde{r}_0=\frac{\sqrt{32\pi P+1}+1}{16\pi P}$. When the black hole evaporates and loses mass, the black hole temperature decreases so the evaporation process is increasingly difficult. Near the extremal black hole the evaporation slows down quickly, and it will take infinite time to reach it, which obeys the third law of thermodynamics.

For $c_0>0$, the black hole is in an analogy with the Hawking-Page phase transition in Schwarzschild AdS black hole. When $r_0\rightarrow 0$, $T\rightarrow \infty$ in all three cases. The black hole can always evaporate away in a finite time. In the cases of $c_1=0$ and $c_1<0$, the black hole mass is unbounded. When the initial mass is taken to infinity, we can observe the decay time is also finite and is in the \emph{linear} order of $\ell$, which is different from the $\ell^3$ in the Einstein gravity. This is due to the different formulas of the black hole mass in two gravity theories.

Similar analysis may also be extended to charged and/or rotating black holes. In Einstein gravity, Page proved an asymptotically flat rotating neutral black hole always loses angular momentum more rapidly than it loses mass \cite{Page:1976ki}. For the charged black hole, the production of charged particles should be treated separately from the massless particles, but they are all part of Hawking emission. The explicit evolution process depends on both the initial black hole mass and charge \cite{Hiscock:1990ex,Ong:2014maa,Ong:2014nha}. It would be interesting to consider such examples in conformal gravity. Furthermore, in the present work we only concentrate on the qualitative features of the black hole evaporation. To get a precise lifetime one has to compute the grey body factors of all the particles in the theory \cite{Gubser:1997yh,Gubser:1997qr,Mathur:1997et,Klemm:1998bb,Kanti:2002nr,Kanti:2002ge,
Harris:2003eg,Kanti:2004nr,Kanti:2005ja,Cardoso:2005vb,Cardoso:2005mh,Harmark:2007jy,Boonserm:2008zg,Chen:2010ru,Miao:2017jtr,
Zhang:2017yfu}. More importantly, near the end of the evaporation, the black hole gets extremely hot and effective field theory starts to fail. New physics, maybe quantum gravity, could affect the subsequent evolution. See e.g. \cite{Adler:2001vs,Chen:2014jwq,Ong:2018syk,Yao:2018ceg} for discussion and references in this field. This process has yet to be fully understood. We hope to be able to consider these questions in our future work.

\begin{acknowledgments}

We acknowledge the  support by Natural Science Foundation of Guangdong Province (2017B030308003) and the Guangdong Innovative and Entrepreneurial Research Team Program (No.2016ZT06D348), and the Science Technology and Innovation Commission of Shenzhen Municipality (ZDSYS20170303165926217, JCYJ20170412152620376).

\end{acknowledgments}

\providecommand{\href}[2]{#2}\begingroup\raggedright\endgroup

\end{document}